\documentclass[12pt,a4paper]{article}
\usepackage[]{times}
\usepackage{dcolumn}
\newcolumntype{d}[1]{D{.}{.}{#1}}
\usepackage[dvips]{epsfig}
\usepackage[latin1]{inputenc}
\usepackage[margin=2.5cm]{geometry}
\usepackage[mathscr]{eucal}
\usepackage[section]{algorithm}
\usepackage[sumlimits, intlimits, namelimits]{amsmath}
\usepackage[T1]{fontenc}

\usepackage{algpseudocode,amsfonts,amssymb,amsthm,bbm,booktabs,braket,comment,fancyhdr,float}
\usepackage{geometry,graphicx,multirow,rotating,sistyle,subfigure,url,comment}

\usepackage[pdftex, pdftitle = {Spatially adaptive Bayesian estimation
  for probabilistic temperature forecasts}, pdfauthor = {AM},
  pdfsubject = {Probabilistic weather forecasting}, pdfpagemode =
  UseOutlines, bookmarks, bookmarksopen, bookmarksnumbered,
  hyperindex, hypertexnames, pdfview=Fit, pdfstartview = Fit,
  pdfpagelabels, colorlinks, linkcolor = black, urlcolor = black,
  citecolor = blue]{hyperref}

\def\real{\mathbb{R}}

\def\bs{\mathbf{s}}

\def\bX{\mathbf{X}}
\def\by{\mathbf{y}}
\def\cN{{\cal N}}
\def\dd{\mathrm{d}}
\def\myE{{\ensuremath{\mathbb{E}}}}
\def\one{\ensuremath{\mathbbm{1}}}

\usepackage{natbib}
\begin{document}

\title{Spatially adaptive, Bayesian estimation for probabilistic temperature forecasts}

\author{Annette M\"{o}ller \\ {\em University of G\"ottingen} \\
        Thordis L.~Thorarinsdottir, Alex Lenkoski \rule{0mm}{7mm} \\
        {\em Norwegian Computing Center, Oslo} \\
        Tilmann Gneiting \rule{0mm}{7mm} \\
        {\em Heidelberg Institute for Theoretical Studies and Karlsruhe Institute of Technology}}

\maketitle

\begin{abstract}
\noindent
Uncertainty in the prediction of future weather is commonly assessed
through the use of forecast ensembles that employ a numerical weather
prediction model in distinct variants.  Statistical postprocessing can
correct for biases in the numerical model and improves calibration.
We propose a Bayesian version of the standard ensemble model output
statistics (EMOS) postprocessing method, in which spatially varying
bias coefficients are interpreted as realizations of Gaussian Markov
random fields.  Our Markovian EMOS (MEMOS) technique utilizes the
recently developed stochastic partial differential equation (SPDE) and
integrated nested Laplace approximation (INLA) methods for
computationally efficient inference.  The MEMOS approach shows
good  predictive performance in a comparative study of 24-hour
ahead temperature forecasts over Germany based on the 50-member
ensemble of the European Centre for Medium-Range Weather Forecasting
(ECMWF).

\end{abstract}

\section{Introduction}

The two major sources of uncertainty in numerical weather prediction
lie in the formulation of the physics based model and in the choice of
initial and boundary conditions.  These are commonly addressed by
using ensemble prediction systems that generate probabilistic forecast
ensembles, where the members vary in the details of the numerical
model and/or initial and boundary conditions
\citep{GneitingRaftery2005, LeutbecherPalmer2008}.

Statistical postprocessing is employed to correct for biases and
dispersion errors in forecast ensembles, based on training data from
the past.  Specifically, ensemble model output statistics (EMOS) and
ensemble Bayesian model averaging (BMA) are widely used postprocessing
techniques that yield full predictive distributions from ensemble
output; for recent reviews, see, e.g., \citet{WilksHamill2007},
\citet{Schefzik&2013}, and \citet{GneitingKatzfuss2014}.  Ensemble BMA
assigns a kernel density to each bias-corrected ensemble member and
combines them into a mixture distribution, using weights that reflect
the skill of the individual members.  In contrast, EMOS uses a
parsimonious distributional regression framework, where a single
parametric predictive distribution is obtained, with the parameters
depending on the ensemble members in suitable ways.
\citet{Raftery&2005} and \citet{Gneiting&2005} developed these methods
for forecasts of temperature and pressure, with the Gaussian model
supplying the underlying kernel density and parametric predictive
distribution, respectively.  For other weather quantities, alternative
distributions are required, as discussed by \citet{Schefzik&2013} and
the references therein.

The main objective of statistical postprocessing is to correct for
systematic shortcomings in the numerical model.  These biases may
differ from location to location, due to, e.g., incomplete resolution
of the orography or land use characteristics by the model grid.
Similarly, the prediction uncertainty may vary over space in ways not
represented by the ensemble spread.  Consequently, the initial Global
EMOS approach \citep{Raftery&2005, Gneiting&2005}, which uses
parameters that are spatially constant, entails lesser predictive
performance than a Local EMOS approach
\citep{ThorarinsdottirGneiting2010}, which estimates a predictive
model at any observation station, based on training data at the
station alone.  However, the Global approach allows for predictions at
just any location, and this is important, as there is an acute need
for statistical postprocessing on gridded domains \citep{MassETAL2008,
GlahnETAL2009}.  In this light, \citet{Kleiber&a2011, Kleiber&b2011}
developed a geostatistical approach that estimates ensemble BMA parameters at
each location separately and, subsequently, interpolates them to
arbitrary locations by employing a spatial statistical model.
\citet{ScheuererBuermann2014} and \citet{ScheuererKoenig2014}
developed a locally adaptive EMOS approach, where information about
the short-term local climatology is incorporated and an approach based
on intrinsic Gaussian random fields is used to interpolate away from
observation locations.

In this paper we propose a Bayesian, locally adaptive implementation
of the EMOS method where the regression coefficients for the mean of
the predictive distribution vary in space according to Gaussian random
fields (GRFs).  In order to exploit the most recent available training
data, inference for the postprocessing parameters must be repeated in
every time step, and thus needs to be computationally efficient.  To
this end, \citet{LindRue} provide an explicit Markov random field
representation of GRFs by solving a certain stochastic partial
differential equation (SPDE) on a discretized domain.  We utilize this
approach in concert with the integrated nested Laplace approximation
(INLA) technique \citep{RueMartChop} to efficiently fit our model in a
Bayesian fashion, thereby taking account of estimation uncertainty.
Due to the Markovian structure of the GRF, we call our new method
Markovian EMOS (MEMOS).  While Bayesian inference has been
investigated before for BMA \citep{VrugtDiksClark2008,
DiNarzoCocchi2010}, the MEMOS method provides a Bayesian approach to
EMOS, and the first Bayesian approach to spatially adaptive
postprocessing in general, to our knowledge.

The paper is organized as follows.  Section \ref{sec:methods}
introduces our spatially adaptive, Bayesian estimation approach and
its implementation in the SPDE-INLA framework.  A case study on
forecasts of surface temperature over Germany based on the 50-member
ensemble of the European Centre for Medium-Range Weather Forecasts
(ECMWF) is presented in Section \ref{sec:casestudy}.  We end the paper
in Section \ref{sec:discussion} with a discussion of possible
extensions of our method.

\section{A spatially adaptive, Bayesian approach to inference}  \label{sec:methods}

\subsection{Ensemble model output statistics (EMOS)}  \label{sec:EMOS}

The ensemble model output statistics (EMOS) methodology was introduced
by \citet{Gneiting&2005} as a Gaussian distributional regression
technique.  Given an ensemble forecast $f_1, \ldots, f_m$ for a
univariate quantity $Y$, EMOS employs the ensemble member forecasts as
predictors in the linear model
\[
Y = a + b_1 f_1 + \cdots + b_m f_m + \varepsilon,
\]
where $a, b_1, \ldots, b_m$ are real-valued regression coefficients
and $\varepsilon \sim \cN(0, \sigma^2)$.  For ensembles with
exchangeable members, such as the aforementioned ECMWF ensemble, one
sets $b_1 = \cdots = b_m$, so that the linear model can be written as
\begin{equation}  \label{EMOS}
Y = a + b \bar{f} + \varepsilon,
\end{equation}
where $\bar{f} = \frac{1}{m} \sum_{k=1}^m f_k$ is the ensemble mean
and, again, $\varepsilon \sim \cN(0, \sigma^2)$.  In what follows, we
focus the discussion on this setting.  The parameters $a$, $b$, and
$\sigma^2$ are estimated from forecast and observation data in a
rolling training period.  \cite{Gneiting&2005} compared maximum
likelihood estimation to optimizing the continuous ranked probability
score (CRPS; see eq.~\eqref{CRPS} below) and concluded that the latter
yields superior predictive performance.

In practice, we need to consider predictions over a domain $D$ that
corresponds to the geographic region at hand.  To this end, we write
$\bar{f}(\bs)$ and $Y(\bs)$ to denote the ensemble mean and the
observation at the location $\bs \in D$, respectively.  In the Global
EMOS approach the statistical parameters are held constant across
space, in that
\[
Y(\bs) \, | \, \bar{f}(\bs) \sim \cN \! \left( a + b \bar{f}(\bs), \sigma^2 \right) \! ,
\]
thereby allowing for large, spatially composited training sets, and
yielding estimates with low variance but strong local biases.  In the
Local EMOS technique, the statistical parameters differ from site to
site, so that
\[
Y(\bs) \, | \, \bar{f}(\bs) \sim
\cN \! \left( a(\bs) + b(\bs) \bar{f}(\bs), \sigma^2(\bs) \right) \! ,
\]
where the statistical parameters are estimated from training data at
the location $\bs \in D$ only.  This yields estimates with small local
biases but high variances.  As noted, the Global approach allows for
predictions at just any location, and geostatistical approaches admit
the interpolation of Local EMOS parameters from spatially scattered
stations to just any location.

\subsection{MEMOS}  \label{sec:MEMOS}

Our Markovian EMOS (MEMOS) technique interprets the spatially varying
bias parameters $a(\bs)$ and $b(\bs)$ as realizations of GRFs, using
the SPDE-INLA framework for computationally efficient Bayesian
inference.  This allows the coefficients to adapt to local conditions,
while using training data from all observation sites, thereby
combining the benefits of both the Global and the Local approaches,
and outperforming either.

Specifically, let $\{ Y(\bs) : \bs \in D \}$ denote the future
temperature field over the domain $D$, and let $D_0 \subset D$ denote
the combined set of observation and prediction locations.  The MEMOS
approach utilizes a latent Gaussian Markov random field (GMRF)
representation for the bias parameters $\{ a(\bs) : \bs \in D_0 \}$
and $\{ b(\bs) : \bs \in D_0 \}$ along with Bayesian inference in the
SPDE-INLA framework.  The result is a joint posterior distribution of
the spatially varying bias parameters and the spatially constant
variance term $\sigma^2$, given the training data.  Conditional on the
values of $\bar{f}(\bs)$, $a(\bs)$, $b(\bs)$, and $\sigma^2$, the
predictive distribution at $\bs \in D_0$ is Gaussian,
\[
Y(\bs) \, | \, \bar{f}(\bs), a(\bs), b(\bs), \sigma \sim
\cN \! \left( a(\bs) + b(\bs) \bar{f}(\bs), \sigma^2 \right) \! .
\]
Integrating over the joint posterior of the postprocessing parameters,
we obtain the posterior predictive distribution of the future
temperature as a Gaussian variance-mean mixture \citep{BNKS82}.  We
approximate this posterior by the finite mixture distribution
\begin{equation}  \label{MEMOS.mixture}
\frac{1}{n} \sum_{i=1}^n
\cN \! \left( a_i(\bs) + b_i(\bs) \bar{f}(\bs), \sigma_i^2 \right) \! ,
\end{equation}
where $(a_1(\bs), b_1(\bs), \sigma_1), \ldots, (a_n(\bs), b_n(\bs),
\sigma_n)$ is a sample from the posterior distribution of the random vector
$(a(\bs), b(\bs), \sigma)$.  In practice, we represent the predictive
distribution by a sample of size $N = mn$ from the finite mixture
distribution in \eqref{MEMOS.mixture}, namely
\begin{equation}  \label{MEMOS.sample}
x_{ij}(\bs) = a_i(\bs) + b_i(\bs) \bar{f}(\bs) + \sigma_i z_j,
\qquad i = 1, \ldots, n, \quad j = 1, \ldots, m,
\end{equation}
where $z_j$ denotes the standard normal quantile at level
$(2j-1)/(2m)$.  The advantages of the specific form of the sample in
\eqref{MEMOS.sample} will become apparent in Section \ref{sec:ECC}.
In our case study, we use this approach with $m = 50$ and $n = 100$,
resulting in a sample of size $N = 5000$.

Before we describe our approach to Bayesian inference using SPDE-INLA,
we review some basic facts about GMRFs.  As is well known, a
stationary GRF $\{ X(\bs) : \bs \in D \subseteq \real^d \}$ with
Mat\'{e}rn covariance function \citep{Matern, GuttorpGneiting2006}
is a solution to the linear fractional SPDE
\begin{equation}  \label{SPDE}
(\kappa^2 - \triangle)^{\alpha/2} \, (\tau X(\bs)) = W(\bs),
\end{equation}
where $\triangle$ is the Laplace operator and $W$ is Gaussian white
noise with unit variance.  Here $\kappa > 0$ is a length scale, $\tau
> 0$ is proportional to the marginal standard deviation, and $\alpha =
\nu + \frac{d}{2}$ is a smoothness parameter, where $\nu > 0$.  When
$\alpha$ is an integer the field has the continuous Markov property
\citep[see, e.g.,][]{RueHeld2005} and admits a GMRF representation.
We have tested models with $\alpha = 1$ and $\alpha = 2$, where the
latter corresponds to a smoother field, and have found that $\alpha =
1$ yields better predictive performance, so from now on we fix $\alpha
= 1$.

\citet{LindRue} proposed an approximate solution of the SPDE
\eqref{SPDE} by discretizing the spatial domain and, subsequently,
solving the equation at the nodes of the grid.  As our observations
are irregularly spaced, we employ a mesh with triangular grid cells.
The algorithm for the mesh creation uses the observation locations as
the initial set of nodes, and iteratively adds new nodes until all the
triangles in the mesh fulfill a set of regularity conditions, with an
equilateral triangle being the most regular.  The approximate solution
to the SPDE is then the weighted sum
\begin{equation}  \label{basisfuncrep}
X_K(\bs) = \sum_{k=1}^K w_k \, \psi_k(\bs)
\end{equation}
of continuous, piecewise linear basis functions $\psi_k(\bs)$ with
support on the triangles that are attached to the $k$th node.  The
random variables $w_k$ are jointly normal with a covariance matrix
that depends on the Mat\'{e}rn parameters $\kappa > 0$ and $\tau > 0$.

\subsection{Bayesian inference using SPDE-INLA}  \label{sec:inference}

To perform Bayesian inference we use the SPDE-INLA approach of
\citet{LindRue} and \citet{RueMartChop} as implemented in the {\tt
  R-INLA} package \citep{R2013, Blangiardo&2013, LindRue2015}.  At
each estimation-prediction step, the observation locations in the
training data are merged with the prediction sites in order to
construct the INLA mesh.  The model has the hyperparameter vector
\[
\left( \kappa_a, \tau_a, \kappa_b, \tau_b, \sigma \right) \! ,
\]
where $\kappa_a$ and $\tau_a$, and $\kappa_b$ and $\tau_b$, are the
Mat\'ern parameters for the GMRF representation of $\{ a(\bs) : \bs
\in D_0 \}$, and $\{ b(\bs) : \bs \in D_0 \}$, respectively.  Our
prior assumes that $\log \kappa_a$ and $\log \kappa_b$ are normal with
mean $-0.082$ and variance 1.5, that $\log \tau_a$ and $\log \tau_b$
are normal with mean $-0.878$ and variance 1.5, and that $1/\sigma^2
\sim \textup{Gamma}(1,0.00005)$, with the components being
independent.  The intercept term $a(\bs)$ is estimated in two parts;
the overall mean level is estimated as a fixed effect, with a random
effect describing the local divergence from the mean.

As a result, {\tt R-INLA} generates samples $\{a_i\}_{i=1}^n$,
$\{b_i\}_{i=1}^n$, and $\{\sigma_i\}_{i=1}^n$ from the marginal
posterior distributions of $a(\bs)$, $b(\bs)$, and $\sigma$,
respectively, or linear combinations thereof, given the training data.
In particular, it generates samples from the marginal posterior
distributions of $\sigma$ and $a(\bs)+b(\bs)\bar{f}(\bs)$, where $\bs$
varies over the prediction sites and $\bar{f}$ is the ensemble mean
for the future time point of interest, from which we obtain the
approximate predictive distribution \eqref{MEMOS.mixture} and its
sample version \eqref{MEMOS.sample}.  An important feature of the
SPDE-INLA approach is that the discrete representation
\eqref{basisfuncrep} defines a field at every location $\bs \in D$.
This is a very useful property for statistical postprocessing directly
on the grid on which the underlying numerical model operates, as is
frequently required in practice \citep{MassETAL2008, GlahnETAL2009}.

\subsection{Ensemble copula coupling}  \label{sec:ECC}

It is important to note that the approximate posterior predictive
distribution \eqref{MEMOS.mixture} and its sample version
\eqref{MEMOS.sample} apply at each site $\bs \in D_0$ independently.
Further effort is needed to generate a sample of physically realistic,
spatially coherent, postprocessed forecast fields.  In the traditional
EMOS approach, an attractive option is to fit a geostatistical model
to the residual process $\{ \varepsilon(\bs) : \bs \in D \}$ in the
basic linear model \eqref{EMOS}, as proposed by \citet{Gel&2004} and
recently developed by \cite{Feldmann&2015}.  Empirical copula based
techniques provide attractive, computationally efficient alternatives,
including both ensemble copula coupling (ECC) and the Schaake shuffle
\citep{Schefzik&2013, Wilks2015}.

Here we pursue the latter approach and propose a slight extension of
the ECC technique.  In a nutshell, ECC restores the rank order
structure of the raw ensemble, which results from a sophisticated
model of the physics and chemistry of the atmosphere, so the adoption
of its dependence structure typically leads to an improvement in the
predictive performance.  Specifically, let $f_1(\bs), \ldots,
f_m(\bs)$ denote the raw ensemble forecast, and let $x_1(\bs), \ldots,
x_m(\bs)$ be a sample from the postprocessed predictive distribution
for $Y(\bs)$, where $\bs$ varies over the set $D_1$ of forecast
locations.  The sample can be generated in various ways, and in our
case study we employ the ECC-Q technique of \citet{Schefzik&2013},
which uses equally spaced quantiles of the univariate postprocessed
predictive distributions.  Finally, let $\pi(\bs)$ denote the
permutation of the integers $1, \ldots, m$ that is induced by the
order statistics of $f_1(\bs), \ldots, f_m(\bs)$, with any ties
resolved at random.  The ECC ensemble at $\bs \in D_1$ is then given
by
\[
g_1(\bs) = x_{(\pi(\bs)(1))}(\bs), \quad \ldots, \quad g_m(\bs) = x_{(\pi(\bs)(m))}(\bs).
\]
While at every location $\bs \in D_1$ individually, this is just a
reordering, the ECC forecast fields $\{ g_j(\bs) : \bs \in D_1 \}$,
where $j = 1, \ldots, m$, inherit the spatial rank order structure of
the raw ensemble.  The basic ECC approach thus produces a
postprocessed ensembles of the same size $m$ as the raw ensemble, and
this is what we do for Global EMOS and Local EMOS.  For MEMOS, the
Bayesian nature and the specific structure of the sample in
\eqref{MEMOS.sample} allows us to apply the basic ECC-Q procedure to
each of the $n$ subsamples $x_{i1}(\bs), \ldots, x_{im}(\bs)$
individually.  Merging the $n$ respective ECC-Q samples, we obtain a
postprocessed ECC ensemble of $N = mn$ spatially coherent forecast
fields.  Of course, the raw ensemble is invariant under ECC; it
already possesses the rank order structure employed in the reordering.

For the purpose of comparison, we assess the predictive performance of
the ECC ensembles relative to Independence ensembles, obtained by
random permutations of the ensemble members at each location
separately.

\subsection{Evaluation methods}  \label{sec:evaluation}

The goal of probabilistic forecasting is to maximize the sharpness of
the predictive distributions subject to calibration
\citep{GneitingBalabdaouiRaftery2007}.  Here we assess calibration via
rank histogram and probability integral transform (PIT) histograms,
and we use proper scoring rules as omnibus performance measures.

As our probabilistic forecasts take various forms, it is important to
use evaluation tools that allow for meaningful comparison.  The raw
ensemble forecast is the discrete, empirical distribution associated
with its $m$ members.  The univariate Local and Global EMOS predictive
distributions in \eqref{EMOS} are Gaussian, and the univariate MEMOS
predictive distribution is the empirical distribution of the sample
\eqref{MEMOS.sample} of size $N = mn$.  In the case of multivariate
probabilistic forecasts at several locations simultaneously, the
Local/Global EMOS and MEMOS predictive distributions are discrete,
corresponding to samples of size $m$ and $N$, respectively.

To assess the calibration of a univariate ensemble of size $m$, we use
the verification rank histogram, which plots the frequency of the rank
of the observation, when pooled with the respective ensemble members
\citep[see, e.g.,][]{Wilks2011}.  For Gaussian predictive
distributions, we use the probability integral transform (PIT)
histogram \citep{Dawid1984, GneitingBalabdaouiRaftery2007}.  The PIT
is simply the value of the predictive cumulative distribution function
evaluated at the verifying observation.  To assess the calibration of
the univariate MEMOS ensemble, we compute the verification rank and
normalize to the unit interval.  In the multivariate case, we employ
the multivariate rank histogram described by \citet{Gneiting&2008}.
In all these cases, calibrated forecasts generate statistically
uniform histograms, and deviations from uniformity can be interpreted
diagnostically.  U-shaped histograms indicate underdispersion, inverse
U-shaped histograms suggest overdispersion, and skewed histograms tend
to be associated with biases.

Proper scoring rules provide summary measures of predictive performance that
may address calibration and sharpness simultaneously
\citep{GneitingRaftery2007}.  We take them to be negatively oriented
penalties, i.e., the smaller, the better.  In the case of univariate
predictive distributions, we use the continuous ranked probability
score (CRPS).  Given the predictive cumulative distribution function,
$F$, and the verifying observation, $y$, the CRPS is defined as
\begin{eqnarray}
\textup{CRPS}(F,y)
& = & \int_{-\infty}^\infty \left( F(x) - \one\{x \geq y\} \right)^2 \dd x \nonumber \\
& = & \myE_F |X-y| - \frac{1}{2} \, \myE_F |X-X'|,  \label{CRPS}
\end{eqnarray}
where $\one\{ \cdot \}$ denotes the indicator function and $X$ and $X'$
are independent with distribution $F$ \citep{GneitingRaftery2007}.
Closed form expressions are available when $F$ is parametric, such as
for Gaussian distributions and finite mixtures thereof
\citep{Grimit&2006}.  The representation \eqref{CRPS} is valid when
$F$ has a finite first moment and implies that the CRPS can be
reported in the same unit as the observation.  When $F$ is an
empirical measure, the expectations become discrete sums, as described
by \citet{Gneiting&2008}, and when $F$ is a deterministic forecast,
i.e., a point measure, the CRPS reduces to the familiar absolute error
(AE).  In computing the AE, we follow \citet{Gneiting2011} and use
the Bayes predictor, namely, the median of the respective predictive
distribution, as point forecast.

In the multivariate setting, where the predictive distribution is for
a quantity in $\real^d$, we use the proper energy score (ES), which is
defined as
\begin{equation}  \label{ES}
\textup{ES}(F,\by) =
\myE_F \|\bX - \by\| - \frac{1}{2} \, \myE_F \|\bX - \bX'\|,
\end{equation}
where $\| \cdot \|$ denotes the Euclidean norm and $\bX$ and $\bX'$
are independent random vectors with distribution $F$ and finite first
moment \citep{GneitingRaftery2007}.  This provides a direct
generalization of the CRPS in the form \eqref{CRPS}.  Again, when $F$
is an empirical measure, the expectations become discrete sums.

To compare and rank the various types of forecasts, we find the mean
CRPS, AE, or energy score, respectively.  In the multivariate setting
(Section \ref{sec:multivariate}) we obtain, for each method, a time
series of scores, and for binary key comparisons we apply the
\citet{DieboldMariano1995} procedure to test the null hypothesis of
equal predictive performance in the form of a vanishing expected score
differential.  In the univariate setting (Section
\ref{sec:univariate}) we report mean scores both over time and over
spatially scattered observation locations.  \citet{HeringGenton2011}
proposed a spatial prediction comparison test (SPCT), which is a
modification of the Diebold-Mariano test in the time series setting.
While this is a very general test, the adaption to our setting is not
straightforward, and we leave it to future research.  Here, we apply
the standard Diebold-Mariano test to the time series of the daily
means of the CRPS and the AE, respectively, and furthermore to the
time series of the scores at individual sites.

\section{Temperature forecasts over Germany} \label{sec:casestudy}

We now present the results of a case study.

\subsection{Ensemble forecast and observation data}  \label{sec:data}

\begin{figure}[t]
\centering
\subfigure[]{\includegraphics[height=90mm]{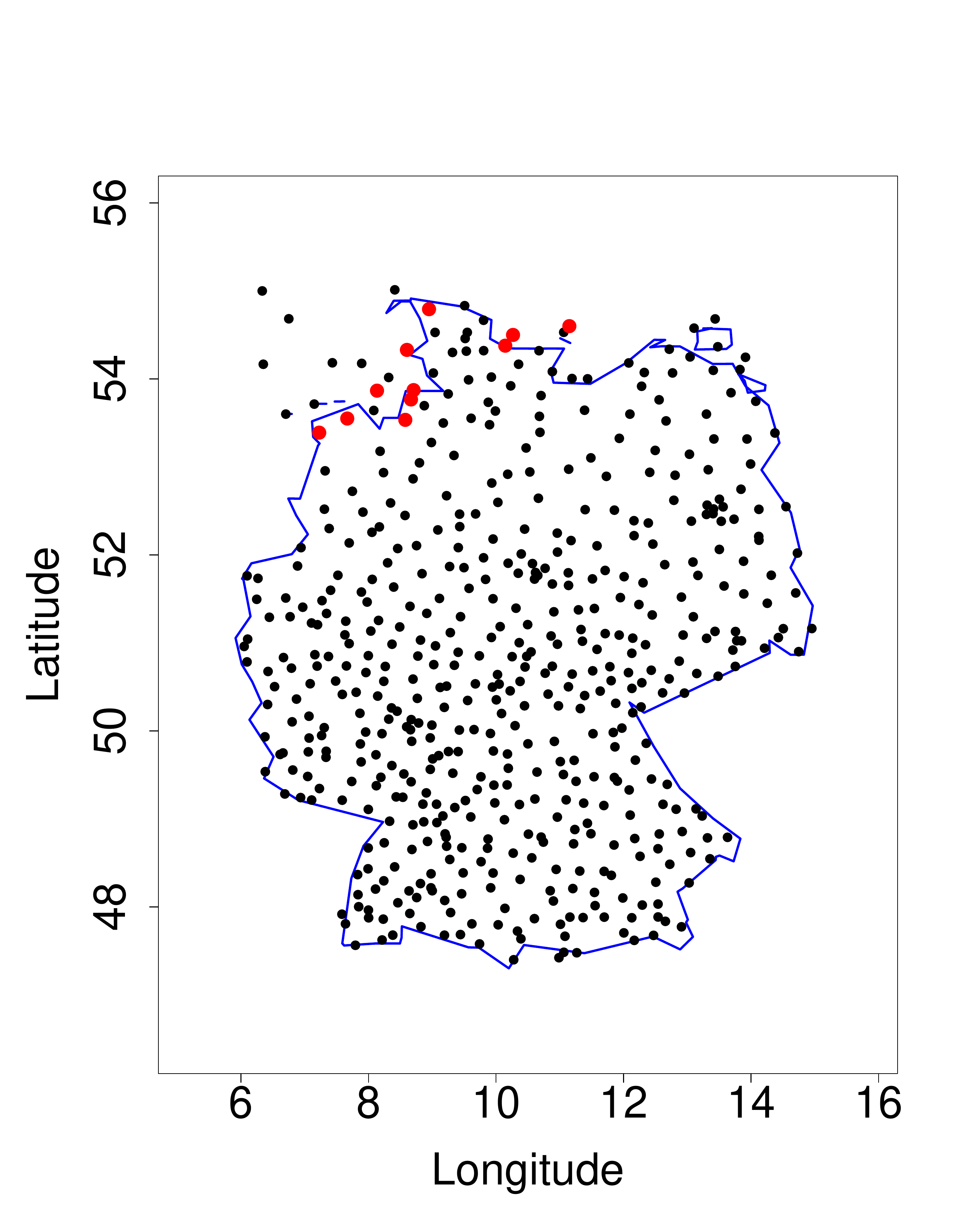}}
\subfigure[]{\includegraphics[height=90mm]{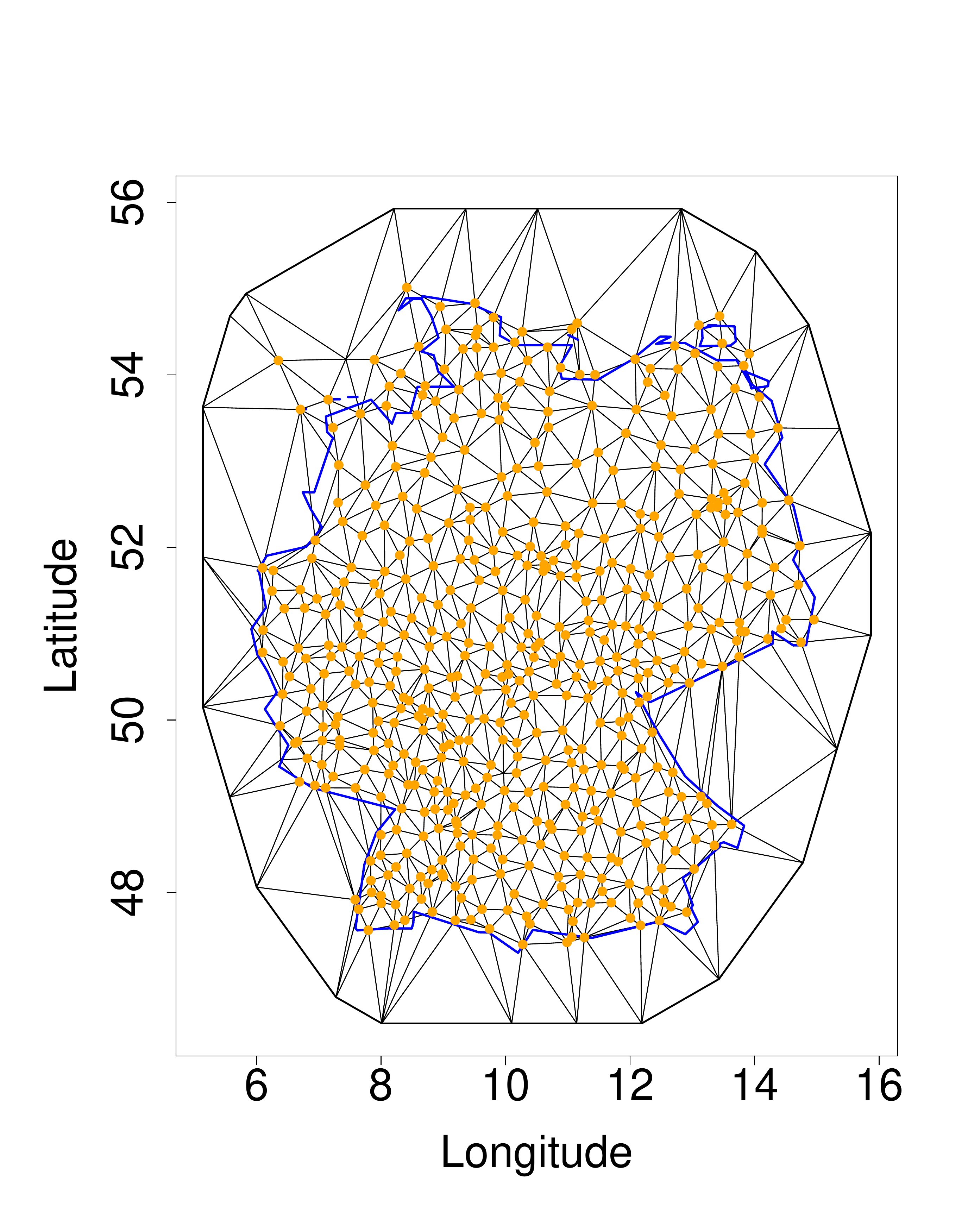}}
\caption{(a) Location of the observation stations over Germany.  The
  eleven stations in the Northern coastline example in Section
  \ref{sec:multivariate} are marked in red.  (b) Triangulation of the
  spatial domain for predictions valid October 3, 2010.
  \label{fig:map}}
\end{figure}

The data in our case study comprises observations and 24-hour ahead
ensemble forecasts of surface (2 meter) temperature over Germany from
February 2, 2010 to April 30, 2011.  The forecast ensemble is the
operational core ensemble of the European Centre for Medium-Range
Weather Forecasts (ECMWF), which has $m = 50$ exchangeable members
\citep{Molteni&1996, Hemri&2014}.  The forecasts are valid at 00
Coordinated Universal Time (UTC), corresponding to 1 am local time in
winter, and 2 am local time during the daylight saving period.
Observations at 518 stations during the considered time period were
obtained and supplied by the German Weather Service (DWD).  As the
ECMWF forecasts are issued on a grid, at 31 km resolution, they are
bilinearly interpolated from the four surrounding grid points to the
station locations of interest.  The unit used is degrees Celsius.

Panel (a) of Figure \ref{fig:map} shows the spatially scattered
locations of the meteorological observation stations in Germany.  The
station density is highest in urban regions, whereas offshore and
outside Germany there are very few stations only.  The triangulation
is flexible, using triangles of different sizes and adapting locally
to the density of the observation locations.  To give an example,
panel (b) shows the mesh created to estimate the MEMOS model from
training data up to October 2, 2010, and used to issue predictions
valid October 3, 2010.  In creating the mesh, we refrain from
extending the data region and require triangles with interior angles
of at least 0.1 degrees.  The
resulting 536 vertices include most, but not all, of the 508 stations
in the training set at hand.

\subsection{Choice of training period}  \label{sec:choice}

There is a tradeoff in the choice of an appropriate rolling training
period for estimating the MEMOS parameters.  Short training periods
adapt quickly to seasonally varying biases; longer training periods
provide more data and reduce the statistical variability in the
estimation.  In the extant literature, rolling training periods of
length between 20 and 40 days have been common practice.  Here we
investigate lengths from 15 to 50 days in five day increments over a
common test period ranging from March 24, 2010 to April 30, 2011.
For MEMOS, both the mean CRPS and the mean AE are
minimal under a training period of 25 days.  For Local EMOS, longer
training periods are slightly favored, but result in negligible
improvement only.  In this light, we use a rolling training period of
length 25 days for all methods.  For Local EMOS, in some instances
observations are available at a subset of the 25 most recent calendar
days only, and then we use the most recent 25 days for which data are
available.  Furthermore, we fix March 24, 2010 to April 30, 2011 as
evaluation period.

\subsection{Results at individual stations}  \label{sec:univariate}

\renewcommand{\arraystretch}{1.1}
\begin{table}[p]
\caption{Mean CRPS and mean AE for raw and postprocessed ensemble
  forecasts of surface temperature at individual stations.  \label{tab:univariate}}
\begin{center}
\begin{tabular}{lcc}
\toprule
            & CRPS & AE \\
\midrule
Raw ECMWF   & 2.50 & 2.81 \\
Global EMOS & 1.79 & 2.49 \\
Local EMOS  & 1.42 & 1.97 \\
MEMOS       & 1.40 & 1.97 \\
\bottomrule
\end{tabular}
\end{center}
\end{table}

\begin{figure}[p]
\centering
\subfigure[Raw ECMWF]{\includegraphics[width=5.5cm]{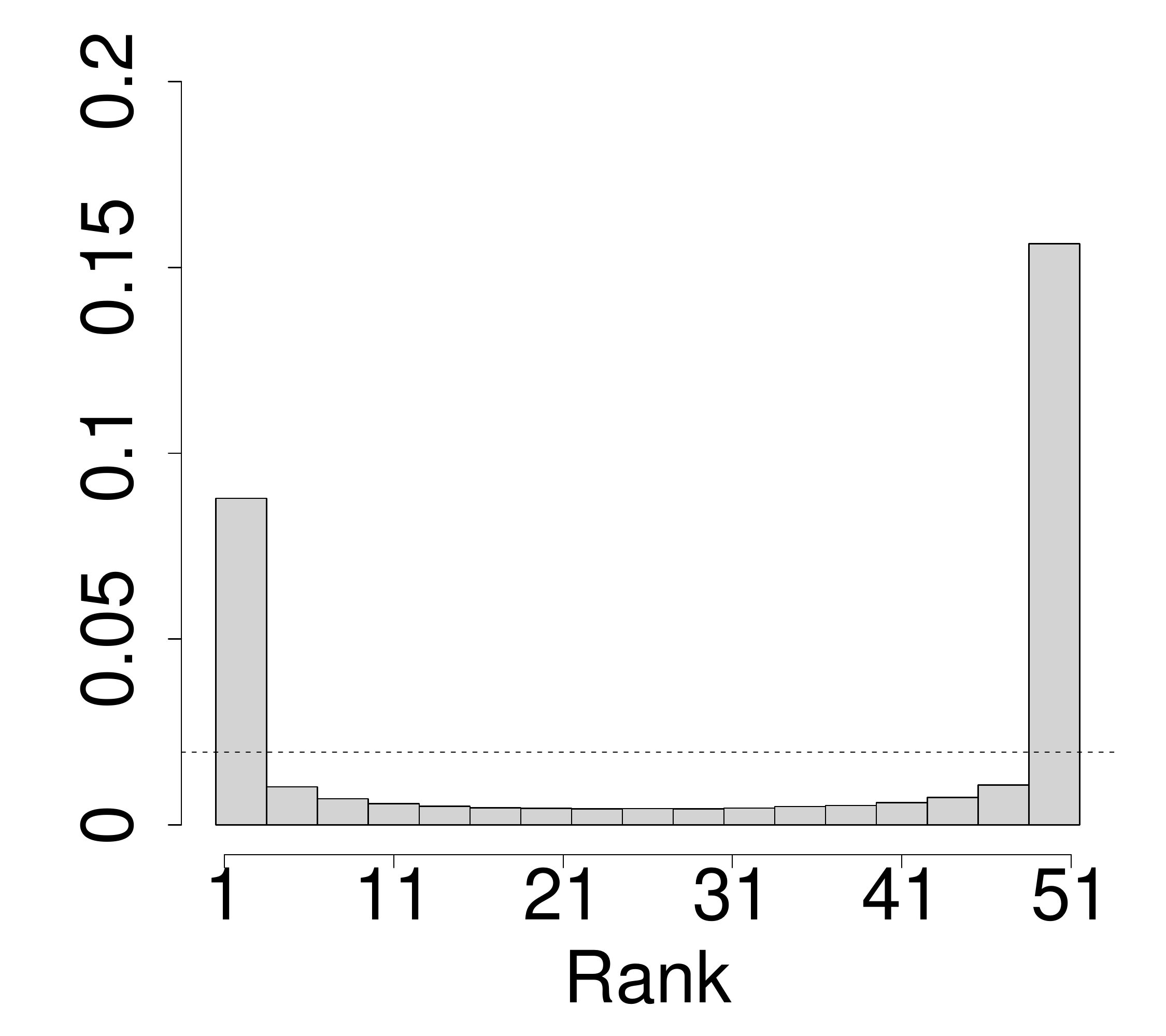}}
\subfigure[Global EMOS]{\includegraphics[width=5.5cm]{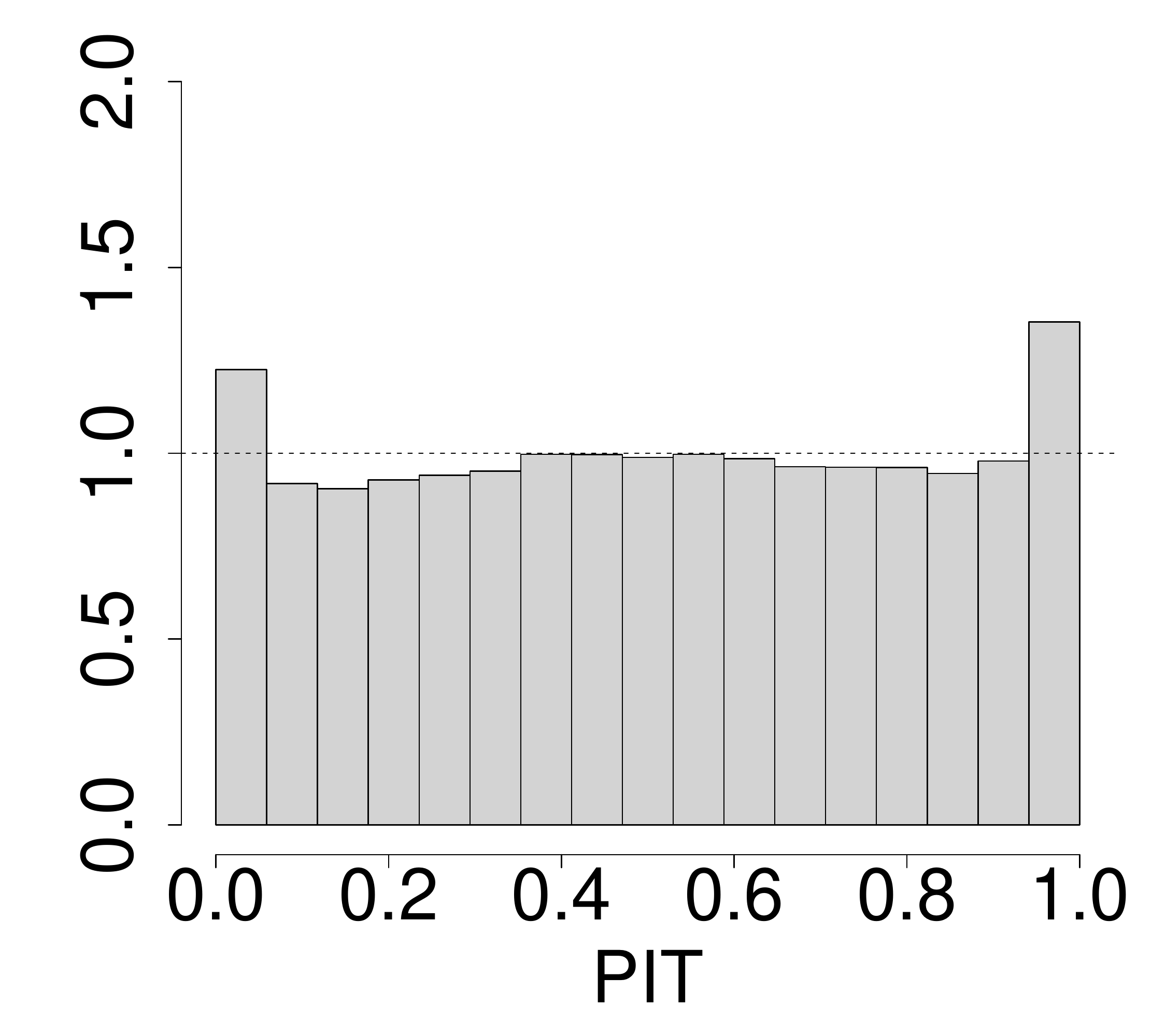}} \\
\subfigure[Local EMOS]{\includegraphics[width=5.5cm]{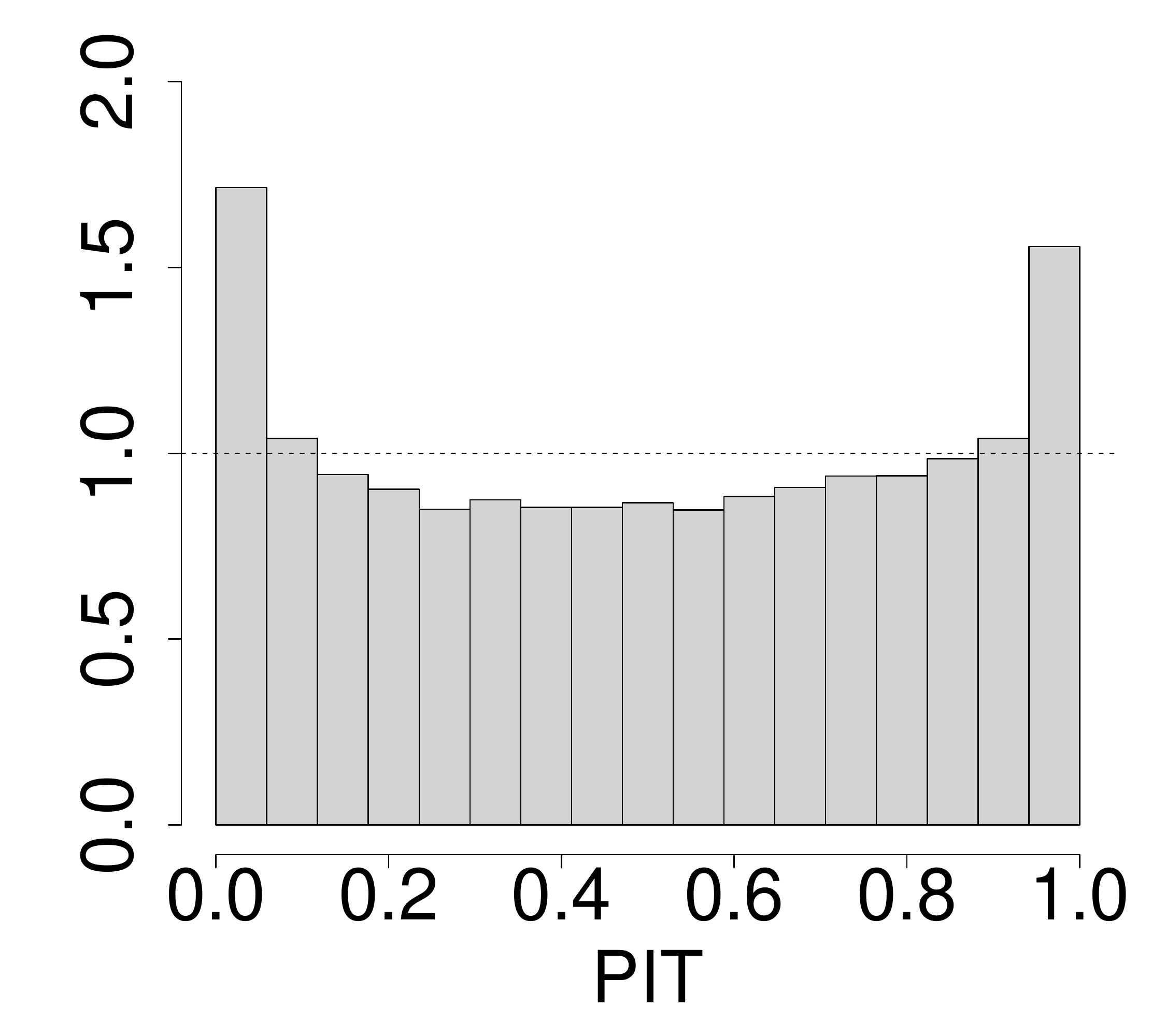}}
\subfigure[MEMOS]{\includegraphics[width=5.5cm]{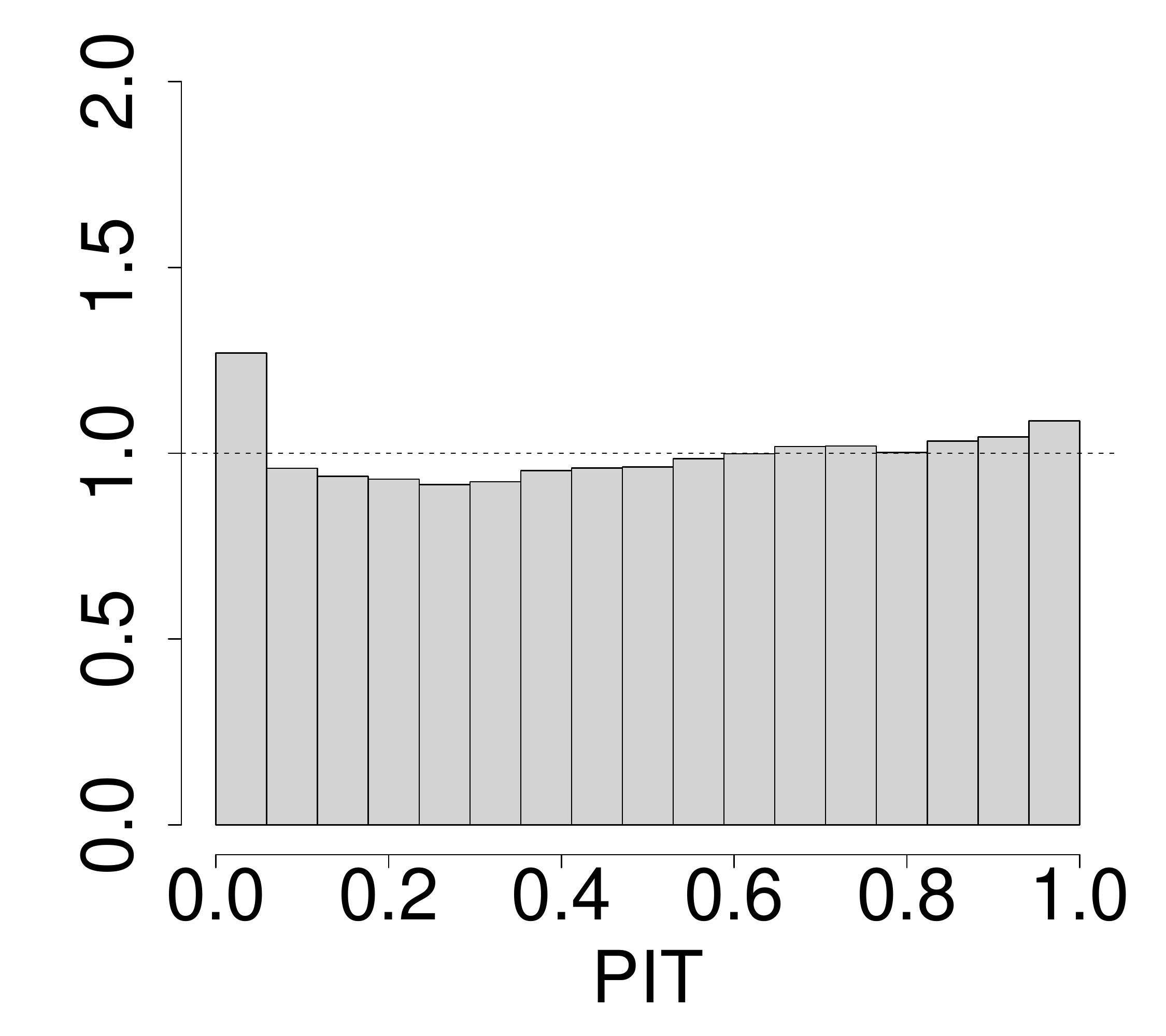}}
\caption{Verification rank and PIT histograms for raw and postprocessed ensemble forecasts
  of surface temperature at individual stations.  \label{fig:univariate}}
\end{figure}

We begin with a comparison of the Raw ECMWF to the postprocessed
Global EMOS, Local EMOS, and MEMOS forecasts at individual stations.
Table \ref{tab:univariate} shows the mean CRPS and mean AE, averaged
over the evaluation period and all available sites, for a total of
205,572 forecast cases. The postprocessed Global
EMOS forecast improves substantially on the Raw ECMWF ensemble, as it
corrects for biases and dispersion errors.  While Global EMOS
estimates only a single set of parameters for all stations, Local EMOS
estimates a separate set of parameters at each considered station, and
Local EMOS improves considerably on Global EMOS.  MEMOS borrows
information from neighbouring stations via the Markovian dependence
structure of the GMRF approximations for the bias parameters, and the
Bayesian implementation takes account of forecast uncertainty, which
leads to further improvement in the predictive performance when measured by the CRPS.

Figure \ref{fig:univariate} shows the rank histogram for the Raw ECMWF
ensemble along with the PIT histograms for Global EMOS, Local EMOS,
and MEMOS, respectively.  Instead of plotting all possible $m + 1 =
51$ bins in the histograms, we use a slightly lower resolution and
aggregate consecutive ranks into 17 bins.  The same resolution is
applied to construct the PIT histograms, for better comparison.  The
rank histogram indicates heavy underdispersion of the Raw ECMWF
ensemble and suggests a pronounced need for postprocessing.  The PIT
histograms for Local EMOS and Global EMOS are much improved, even
though they remain indicative of underdispersion.  MEMOS shows the
most uniform PIT histogram, well in line with the ranking in terms of
the mean CRPS in Table \ref{tab:univariate}.

We now apply the Diebold-Mariano test to the key comparison between
MEMOS and Local EMOS.  For the time series of the daily
mean of the CRPS, the tail probability is smaller than 0.01.  However,
for the daily means of the AE, the tail probability is 0.42 which is
in line with the overall mean AE for MEMOS being equal to that for
Local EMOS, see Table \ref{tab:univariate}.

\subsection{Results at several stations simultaneously}
  \label{sec:multivariate}

\begin{figure}[p]
\centering
\subfigure[Raw ECMWF Independence]{\includegraphics[width=5.3cm, height=4.1cm]{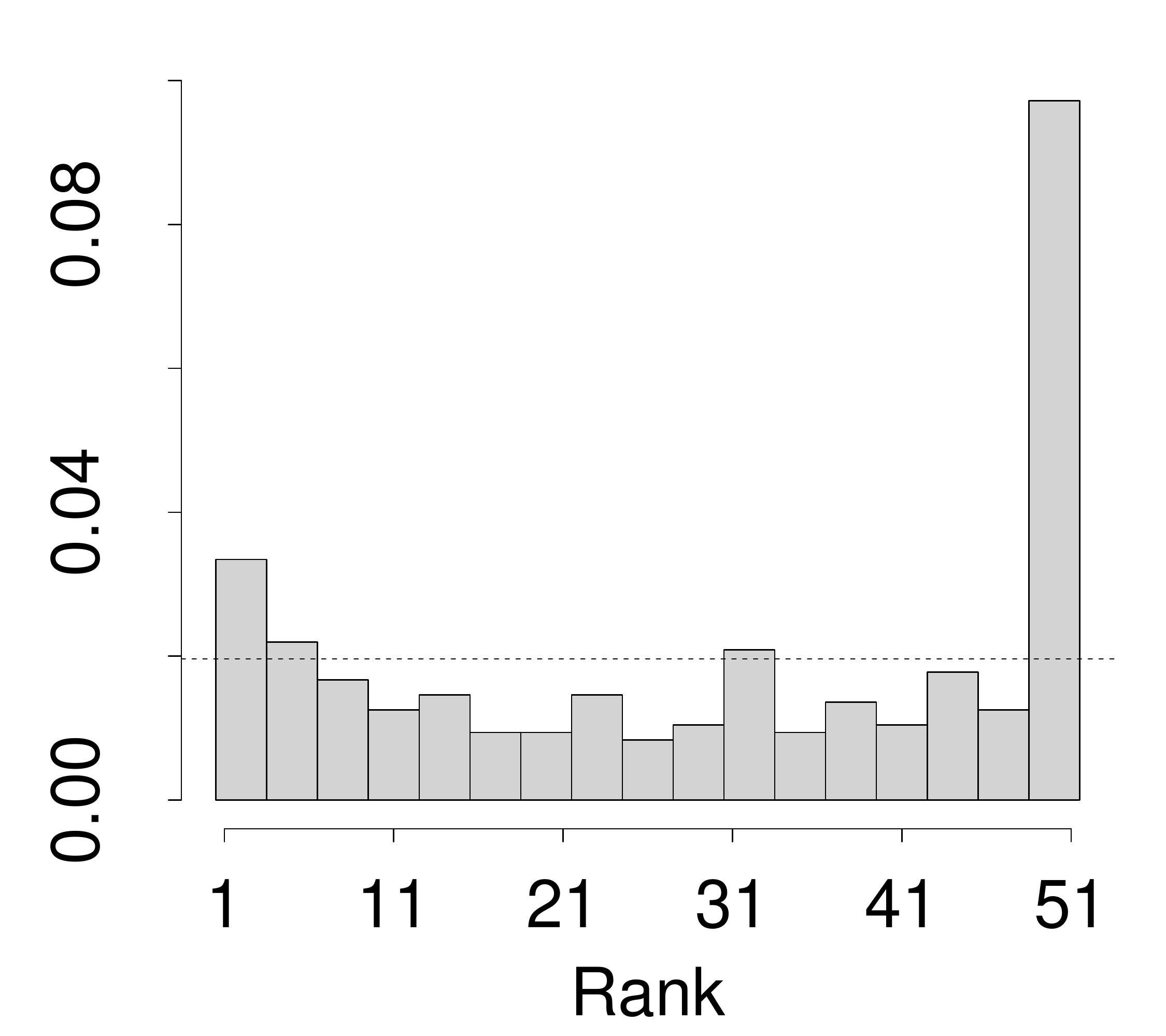}}
\subfigure[Raw ECMWF ECC]{\includegraphics[width=5.3cm, height=4.1cm]{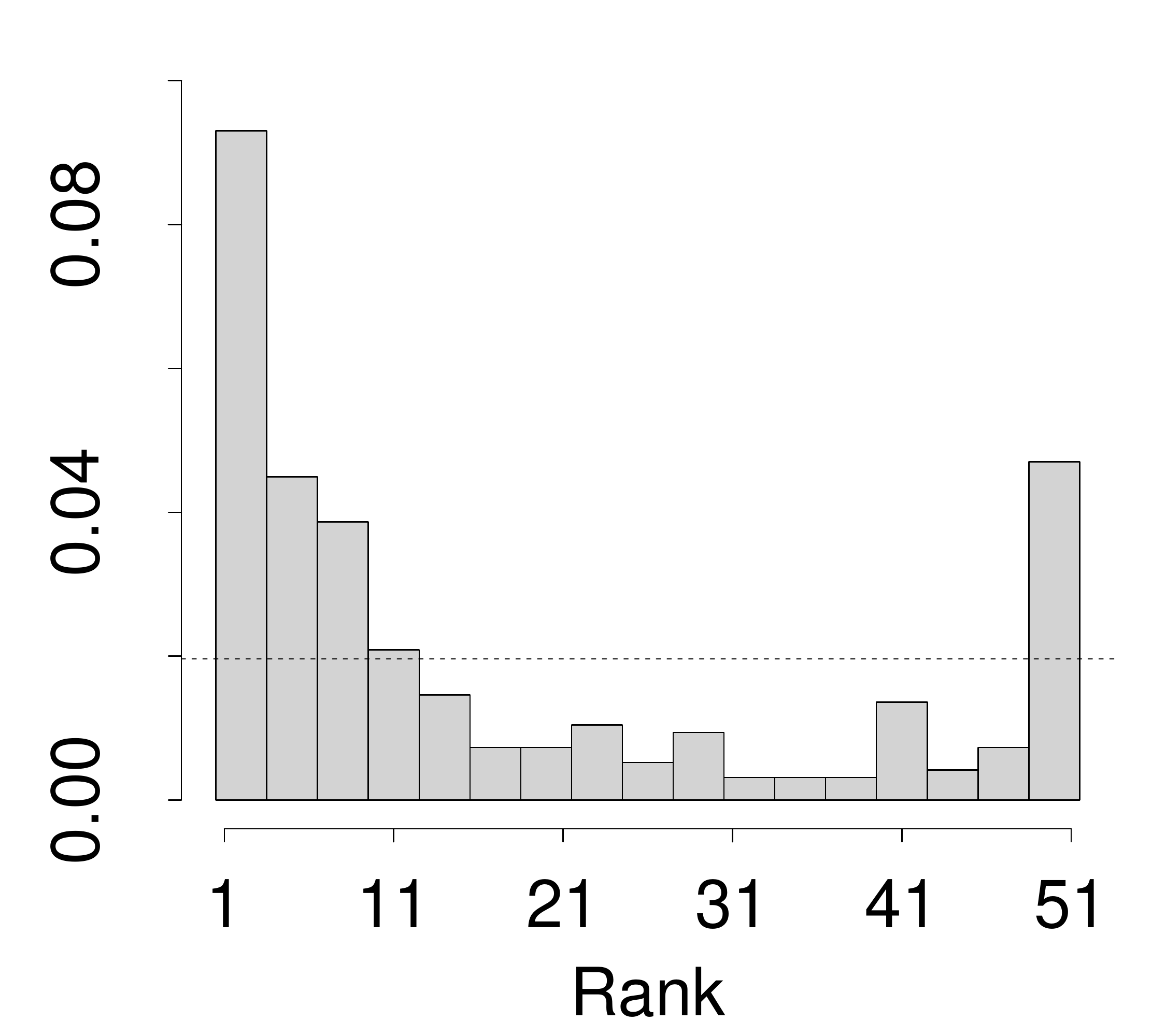}} \\
\subfigure[Global EMOS Independence]{\includegraphics[width=5.3cm, height=4.1cm]{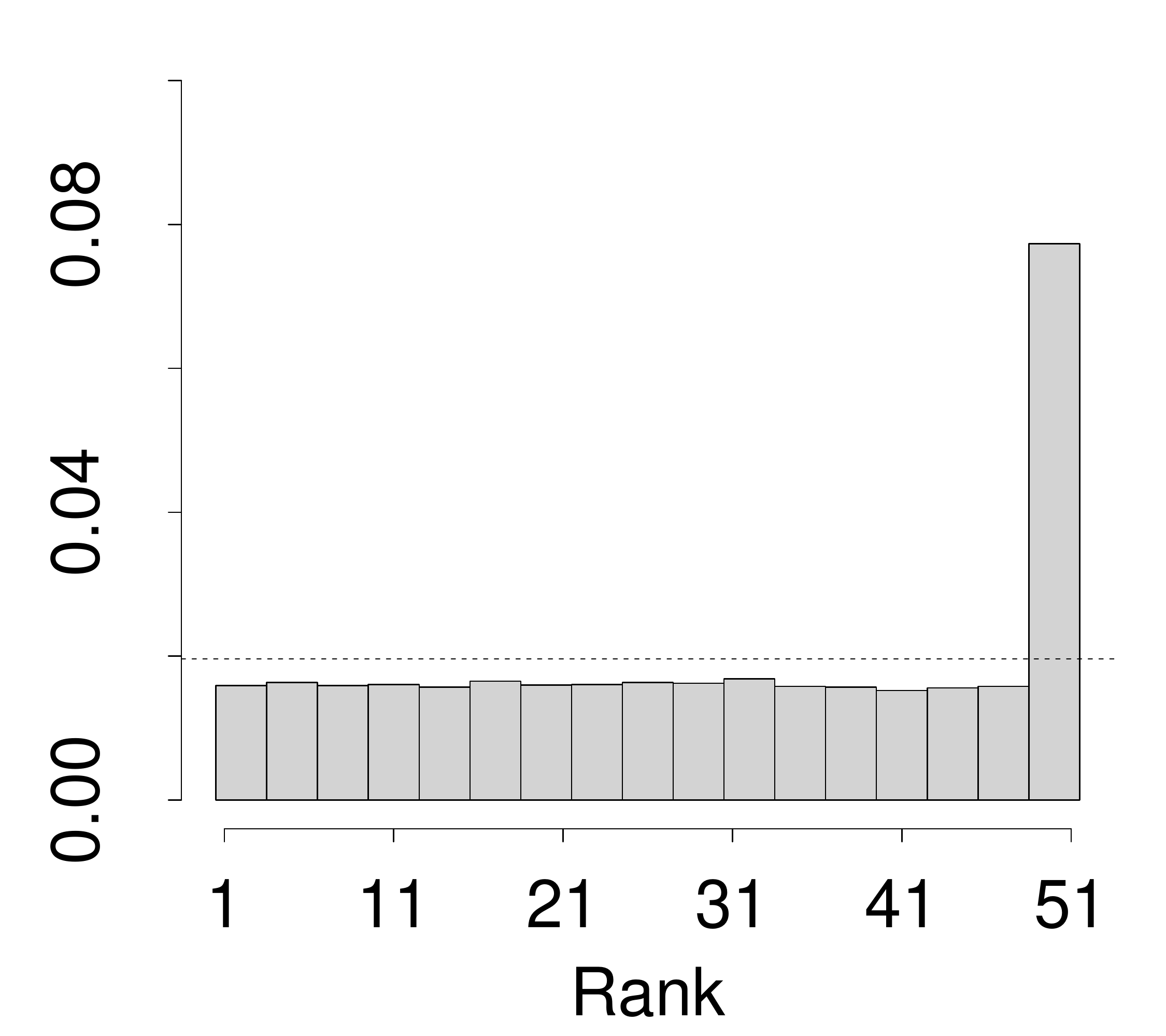}}
\subfigure[Global EMOS ECC]{\includegraphics[width=5.3cm, height=4.1cm]{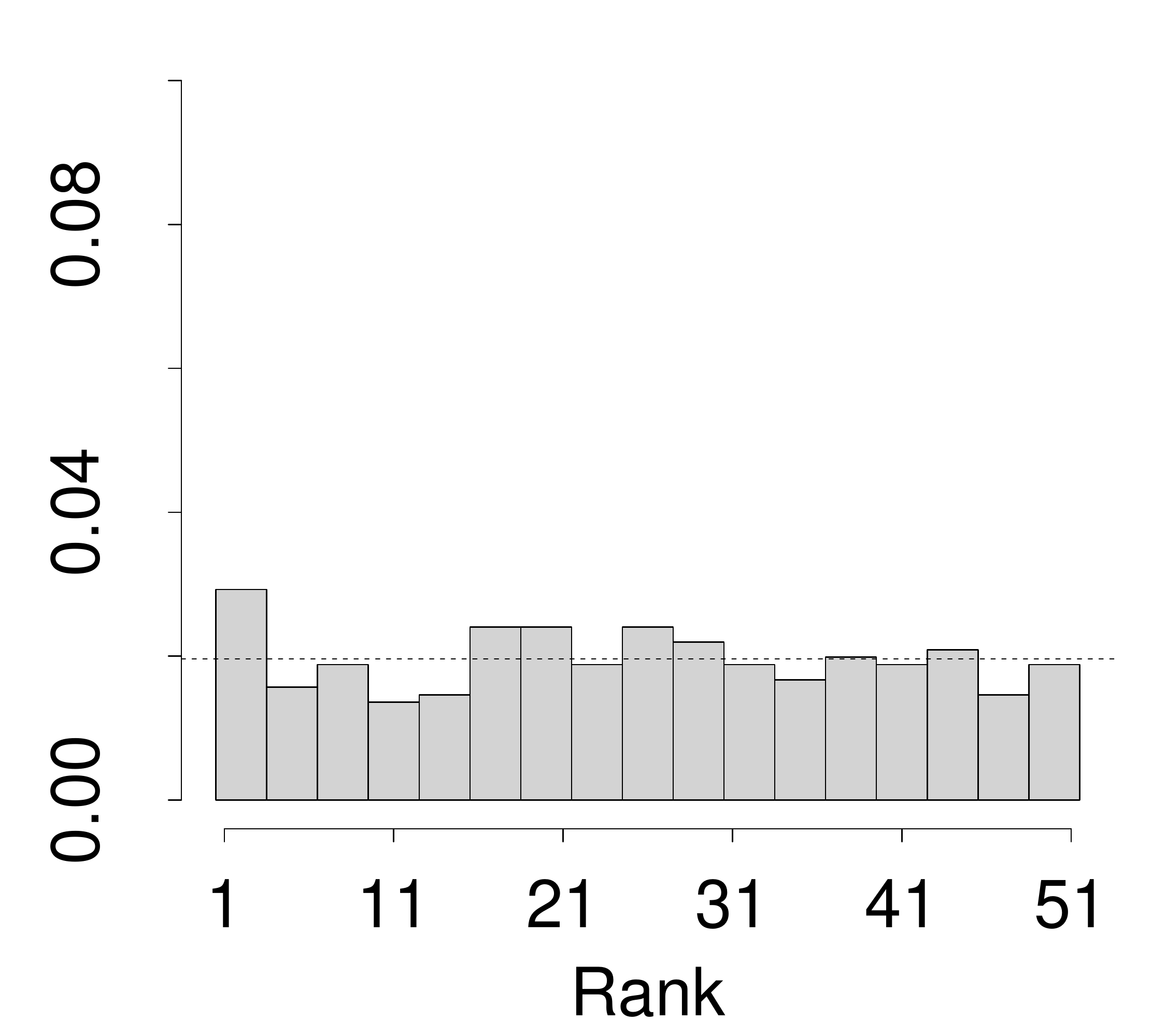}} \\
\subfigure[Local EMOS Independence]{\includegraphics[width=5.3cm, height=4.1cm]{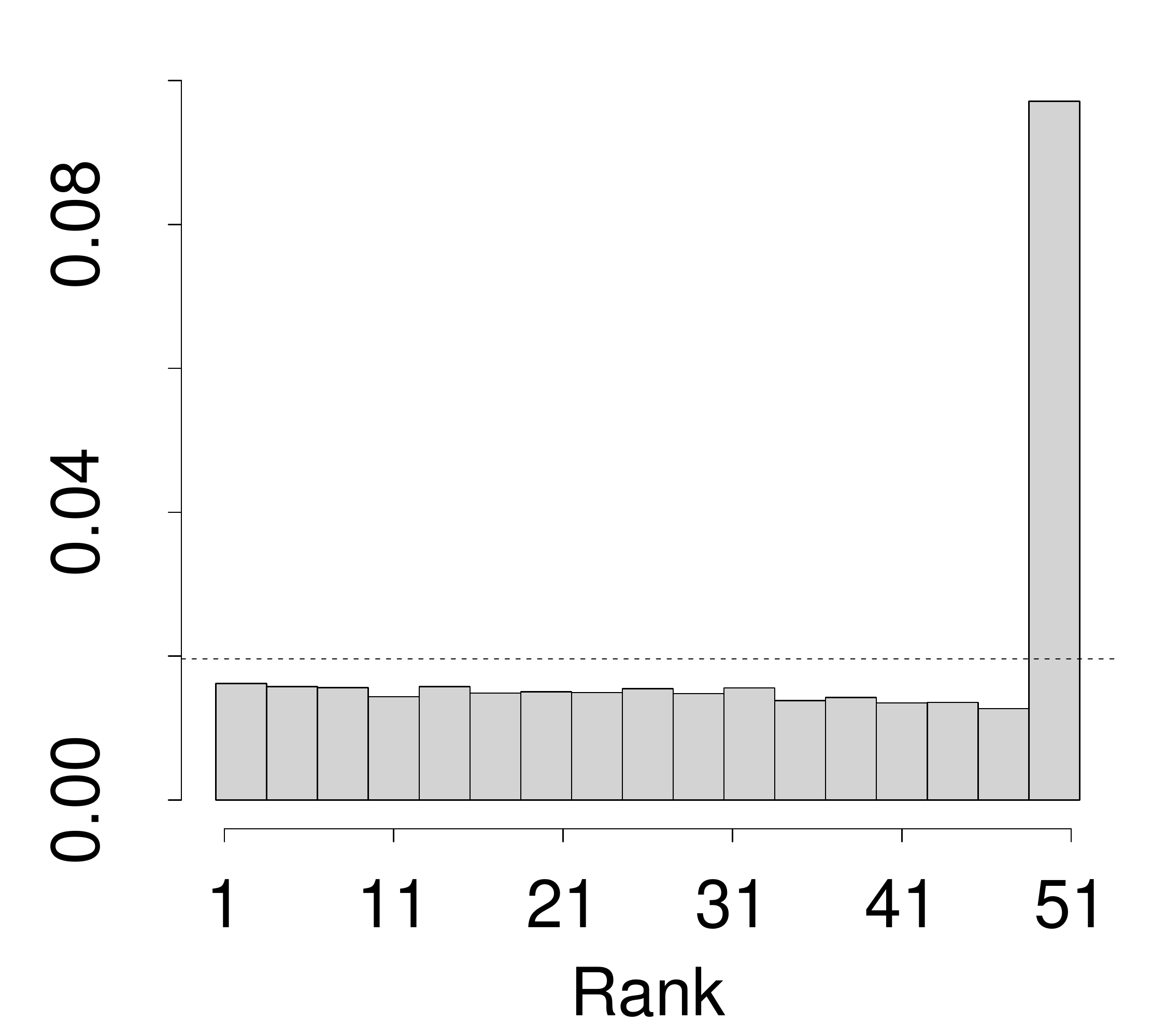}}
\subfigure[Local EMOS ECC]{\includegraphics[width=5.3cm, height=4.1cm]{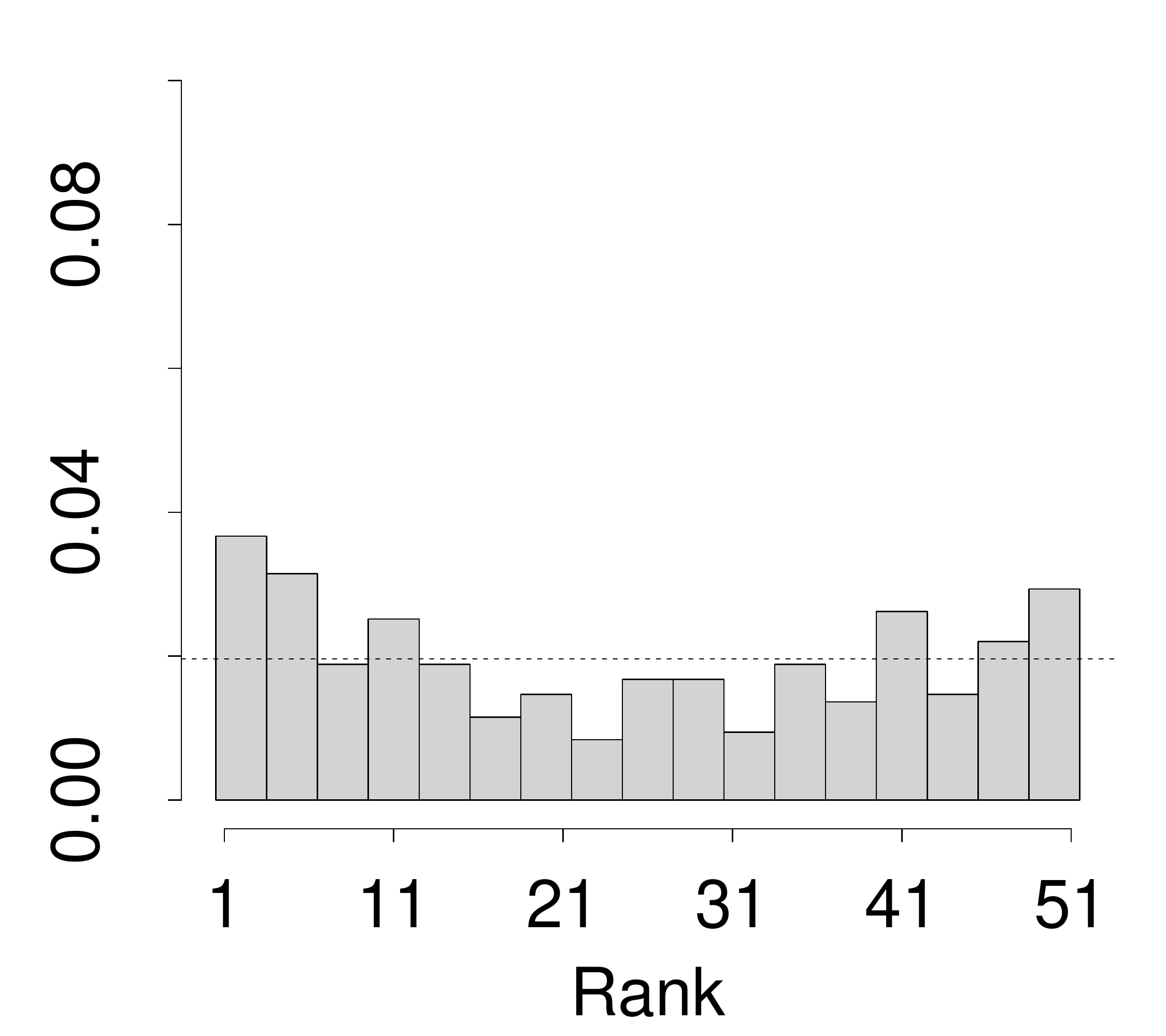}} \\
\subfigure[MEMOS Independence]{\includegraphics[width=5.3cm, height=4.1cm]{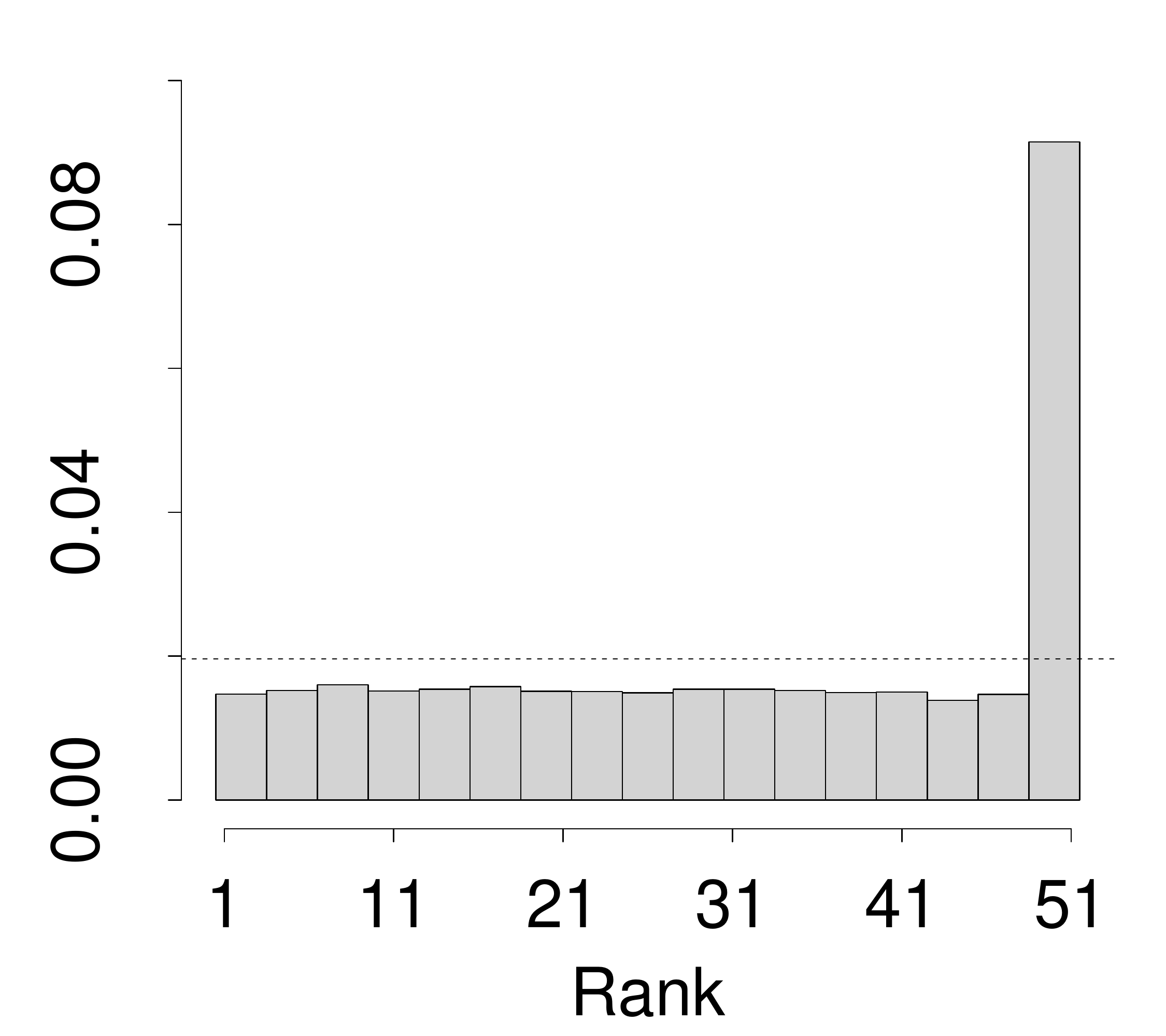}}
\subfigure[MEMOS ECC]{\includegraphics[width=5.3cm, height=4.1cm]{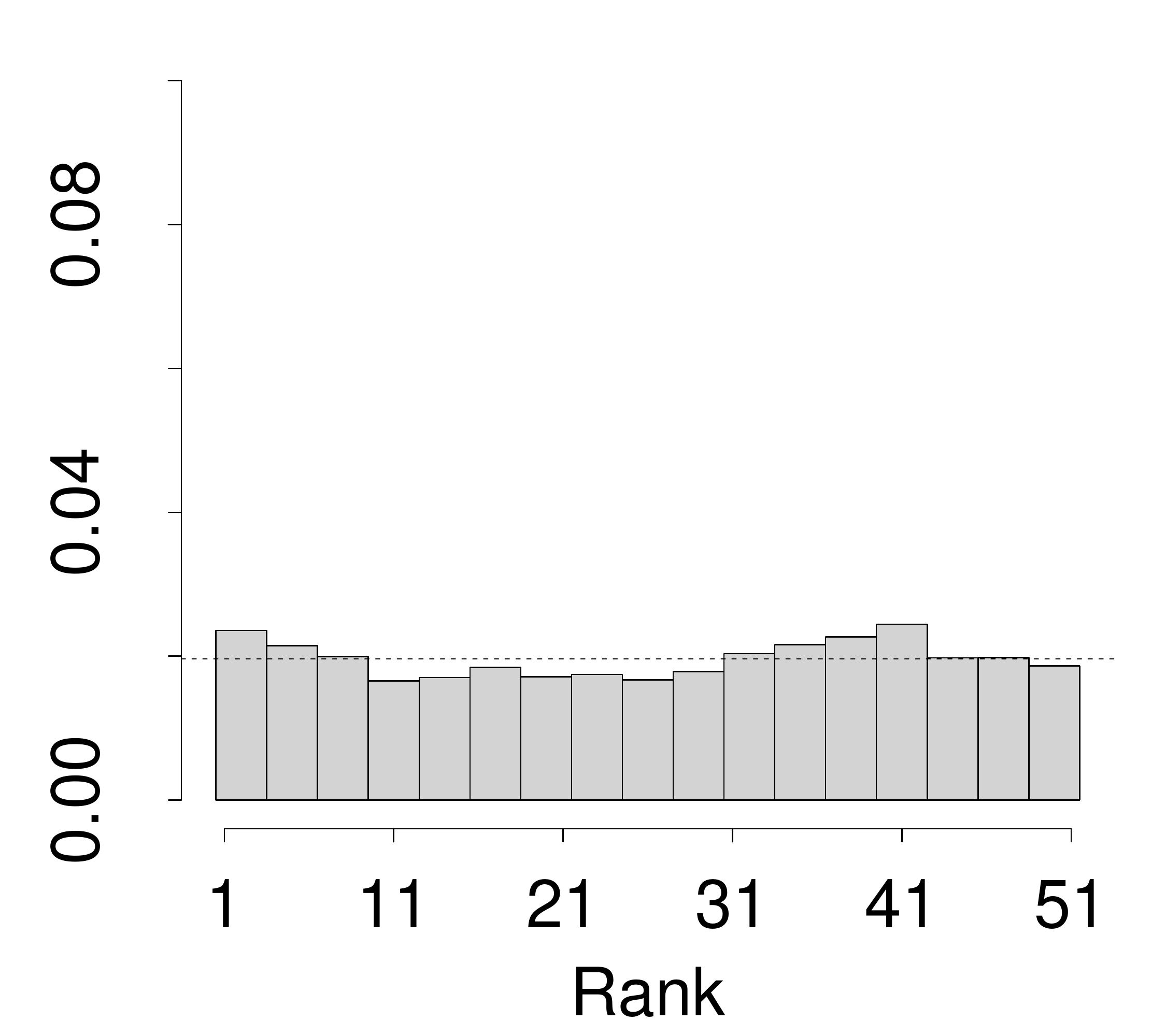}}
\caption{Multivariate rank histograms in the Northern coastline example.  \label{fig:NorthSea}}
\end{figure}

We turn to a multivariate example with eleven stations along the North
Sea and Baltic Sea coastlines, the locations of which are illustrated
in Figure \ref{fig:map}, to compare the predictive performance of the
Raw ECMWF ensemble and the postprocessed Global EMOS, Local EMOS, and
MEMOS forecasts under both ECC and Independence structures.  As noted,
the Raw ECMWF ensemble is invariant under ECC.  The multivariate Raw
ECMWF, Global EMOS, and Local EMOS ensembles are of size $m = 50$,
while the MEMOS ensemble is of size $N = mn = 5,000$, both under ECC
and Independence structures, as described in Section \ref{sec:ECC}.
Table \ref{tab:NorthSea} shows the mean energy score \eqref{ES} and
Figure \ref{fig:NorthSea} the multivariate rank histograms in this
example.  Clearly, ECC improves on the Independence approach, except
in the case of the Raw ECMWF ensemble, where the Independence
assumption compensates for the severe underdispersion, essentially
replacing one evil by another.  Global EMOS improves on the Raw ECMWF
forecast, Local EMOS outperforms Global EMOS, and MEMOS improves on
the Local EMOS benchmark, under the ECC
approach.  For each binary comparison, the tail probability under
the Diebold-Mariano test of equal predictive performance is $\leq
0.01$, except for the comparison between Local EMOS Independence and
MEMOS Independence, where it is 0.31.  Similar patterns can be seen in
the multivariate rank histograms, where Global EMOS ECC and MEMOS ECC
show the most nearly uniform histograms.

\renewcommand{\arraystretch}{1.1}
\begin{table}[t]
\caption{Mean energy score in the Northern coastline example.
  \label{tab:NorthSea}}
\begin{center}
\begin{tabular}{lcc}
\toprule
            & Independence & ECC \\
\midrule
Raw ECMWF   & 6.32 & 6.37 \\
Global EMOS & 5.40 & 5.24 \\
Local EMOS  & 4.79 & 4.74 \\
MEMOS       & 4.81 & 4.66 \\
\bottomrule
\end{tabular}
\end{center}
\end{table}

\section{Discussion}  \label{sec:discussion}

The MEMOS approach provides a spatially adaptive, Bayesian
implementation of the basic EMOS technique, where spatially varying
bias coefficients are modeled as GMRFs.  Inference is performed in a
computationally efficient fashion within the SPDE-INLA framework of
\citet{RueMartChop} and \citet{LindRue}.  Through the Markovian
dependence structure of the GMRFs, MEMOS borrows information from
neighbouring stations at the estimation stage, and the Bayesian
implementation takes account of estimation uncertainty.  In our case
study of probabilistic temperature forecasts over Germany based on the
ECMWF ensemble, MEMOS performs well against the Raw ECMWF ensemble,
Global EMOS, and Local EMOS at observation sites.  Furthermore, MEMOS
allows for statistically postprocessed predictive distributions at any
desired location.

We now describe potential future work.  \citet{ScheuererBuermann2014}
and \citet{ScheuererKoenig2014} developed an EMOS version that
incorporates information about the short-term local climatology, and
their idea could be combined with the MEMOS approach.  Future versions
of MEMOS also might allow for spatially varying predictive variances,
as opposed to the current implementation, which assumes a spatially
constant predictive variance.  In such an extension, the variance
$\sigma^2$ can be parameterized as a linear function of the ensemble
variance, in the sense that ${\rm var}(\varepsilon) = c + d S^2$ in
\eqref{EMOS}, where $S^2 = \frac{1}{m-1} \sum_{k=1}^m (f_k -
\bar{f})^2$ is the ensemble variance \citep{Gneiting&2005}.  The
nonhomogeneous error term accounts for the ensemble spread-skill
relationship as well as possible under- or overdispersion
\citep{WhitakerLoughe}.  Furthermore, MEMOS can be tailored to
ensembles with non-exchangeable members, or ensembles with groups of
exchangeable members.  As a caveat, while these are extensions, they
may or may not prove beneficial in forecast mode.

To account for dependencies in forecasts at several locations
simultaneously, we have combined MEMOS with ECC.  As an alternative,
it may be possible to develop a version of the spatial EMOS technique
of \citet{Feldmann&2015} that applies to MEMOS.  More generally,
methods that account for spatial and temporal as well as
inter-variable dependencies are in high demand.  Recent developments
in these directions include both parametric and non-parametric copula
approaches \citep{MoellerLenkTho2013, Schefzik&2013, Wilks2015,
Schefzik2015}.

Our version of MEMOS is tailored to weather quantities for which the
conditional predictive distributions can be assumed to be Gaussian.
As the SPDE-INLA methodology is not restricted to normally distributed
responses, it may be possible to develop variants for other weather
quantities, utilizing the respective univariate postprocessing
approaches, such as the EMOS techniques proposed for wind speed
\citep{ThorarinsdottirGneiting2010} and precipitation
\citep{Scheuerer2014}, respectively.

As ensemble forecasts continue to improve over time,
\citet{Hemri&2014} analyze the evolution of the difference in skill
between the Raw ECMWF and statistically postprocessed forecasts for a
time period covering the years 2002 to 2014.  Perhaps surprisingly,
they find that the gap in skill remains almost constant over time.
This suggests that improvements in numerical weather prediction models
themselves, and improvements by statistical postprocessing, are
complementary.  In this light, we anticipate that statistical
postprocessing will continue to yield substantial benefits in weather
prediction for decades to come.

\section*{Acknowledgements}

We thank Kira Feldmann, Roman Schefzik, and Michael Scheuerer for
helpful discussions, and Finn Lindgren for invaluable assistance
regarding the understanding of the SPDE methodology and the use of the
{\tt R-INLA} package.  Furthermore, we are grateful to the
European Centre for Medium-Range Weather Forecasts (ECMWF) and the
German Weather Service (DWD) for providing forecast and observation
data, respectively.  Annette M\"oller and Alex Lenkoski gratefully
acknowledge support by the German Research Foundation (DFG) within the
programme ``Spatio-/Temporal Graphical Models and Applications in
Image Analysis'' (GRK 1653).

\bibliography{MEMOS}

\begin{thebibliography}{43}
\providecommand{\natexlab}[1]{#1}
\providecommand{\url}[1]{\texttt{#1}}
\expandafter\ifx\csname urlstyle\endcsname\relax
  \providecommand{\doi}[1]{doi: #1}\else
  \providecommand{\doi}{doi: \begingroup \urlstyle{rm}\Url}\fi

\bibitem[Barndorff-Nielsen et~al.(1982)Barndorff-Nielsen, Kent, and
  S{\o}rensen]{BNKS82}
O.~Barndorff-Nielsen, J.~Kent, and M.~S{\o}rensen.
\newblock Normal variance-mean mixtures and $z$ distributions.
\newblock \emph{International Statistical Review}, 50:\penalty0 145--159, 1982.

\bibitem[Blangiardo et~al.(2013)Blangiardo, Cameletti, Baio, and
  Rue]{Blangiardo&2013}
M.~Blangiardo, M.~Cameletti, G.~Baio, and H.~Rue.
\newblock Spatial and spatio-temporal models with {R-INLA}.
\newblock \emph{Spatial and Spatio-Temporal Epidemiology}, 4:\penalty0 33--49,
  2013.

\bibitem[Dawid(1984)]{Dawid1984}
A.~P. Dawid.
\newblock Statistical theory: The prequential approach (with discussion).
\newblock \emph{Journal of the Royal Statistical Society Series A},
  147:\penalty0 278--292, 1984.

\bibitem[Di~Narzo and Cocchi(2010)]{DiNarzoCocchi2010}
A.~F. Di~Narzo and D.~Cocchi.
\newblock A {B}ayesian hierarchical approach to ensemble weather forecasting.
\newblock \emph{Journal of the Royal Statistical Society Series C},
  59:\penalty0 405--422, 2010.

\bibitem[Diebold and Mariano(1995)]{DieboldMariano1995}
F.~X. Diebold and R.~S. Mariano.
\newblock Comparing predictive accuarcy.
\newblock \emph{Journal of Business and Economic Statistics}, 13:\penalty0
  253--263, 1995.

\bibitem[Feldmann et~al.(2015)Feldmann, Scheuerer, and
  Thorarinsdottir]{Feldmann&2015}
K.~Feldmann, M.~Scheuerer, and T.~L. Thorarinsdottir.
\newblock Spatial postprocessing of ensemble forecasts for temperature using
  nonhomogeneous {G}aussian regression.
\newblock \emph{Monthly Weather Review}, 143:\penalty0 955--971, 2015.

\bibitem[Gel et~al.(2004)Gel, Raftery, and Gneiting]{Gel&2004}
Y.~Gel, A.~E. Raftery, and T.~Gneiting.
\newblock Calibrated probabilistic mesoscale weather field forecasting: The
  geostatistical output perturbation ({GOP}) method (with discussion and
  rejoinder).
\newblock \emph{Journal of the American Statistical Association}, 99:\penalty0
  575--590, 2004.

\bibitem[Glahn et~al.(2009)Glahn, Gilbert, Cosgrove, Ruth, and
  Sheets]{GlahnETAL2009}
B.~Glahn, K.~Gilbert, R.~Cosgrove, D.~P. Ruth, and K.~Sheets.
\newblock The gridding of {MOS}.
\newblock \emph{Weather and Forecasting}, 24:\penalty0 520--529, 2009.

\bibitem[Gneiting(2011)]{Gneiting2011}
T.~Gneiting.
\newblock Making and evaluating point forecasts.
\newblock \emph{Journal of the American Statistical Association}, 106:\penalty0
  746--762, 2011.

\bibitem[Gneiting and Katzfuss(2014)]{GneitingKatzfuss2014}
T.~Gneiting and M.~Katzfuss.
\newblock Probabilistic forecasting.
\newblock \emph{Annual Review of Statistics and Its Application}, 1:\penalty0
  125--151, 2014.

\bibitem[Gneiting and Raftery(2005)]{GneitingRaftery2005}
T.~Gneiting and A.~E. Raftery.
\newblock Weather forecasting with ensemble methods.
\newblock \emph{Science}, 310:\penalty0 248--249, 2005.

\bibitem[Gneiting and Raftery(2007)]{GneitingRaftery2007}
T.~Gneiting and A.~E. Raftery.
\newblock Strictly proper scoring rules, prediction, and estimation.
\newblock \emph{Journal of the American Statistical Association}, 102:\penalty0
  359--378, 2007.

\bibitem[Gneiting et~al.(2005)Gneiting, Raftery, Westveld, and
  Goldman]{Gneiting&2005}
T.~Gneiting, A.~Raftery, A.~Westveld, and T.~Goldman.
\newblock {C}alibrated probabilistic forecasting using ensemble model output
  statistics and minimum {CRPS} estimation.
\newblock \emph{Monthly Weather Review}, 133:\penalty0 1098--1118, 2005.

\bibitem[Gneiting et~al.(2007)Gneiting, Balabdaoui, and
  Raftery]{GneitingBalabdaouiRaftery2007}
T.~Gneiting, F.~Balabdaoui, and A.~E. Raftery.
\newblock Probabilistic forecasts, calibration and sharpness.
\newblock \emph{Journal of the Royal Statistical Society Series B},
  69:\penalty0 243--268, 2007.

\bibitem[Gneiting et~al.(2008)Gneiting, Stanberry, Grimit, Held, and
  Johnson]{Gneiting&2008}
T.~Gneiting, L.~I. Stanberry, E.~P. Grimit, L.~Held, and N.~A. Johnson.
\newblock Assessing probabilistic forecasts of multivariate quantities, with
  applications to ensemble predictions of surface winds (with discussion and
  rejoinder).
\newblock \emph{Test}, 17:\penalty0 211--264, 2008.

\bibitem[Grimit et~al.(2006)Grimit, Gneiting, Berrocal, and
  Johnson]{Grimit&2006}
E.~P. Grimit, T.~Gneiting, V.~J. Berrocal, and N.~A. Johnson.
\newblock The continuous ranked probability score for circular variables and
  its application to mesoscale forecast ensemble verification.
\newblock \emph{Quarterly Journal of the Royal Meteorological Society},
  132:\penalty0 2925--2942, 2006.

\bibitem[Guttorp and Gneiting(2008)]{GuttorpGneiting2006}
P.~Guttorp and T.~Gneiting.
\newblock Studies in the history of probability and statistics {XLIX}: On the
  {M}at{\'e}rn correlation family.
\newblock \emph{Biometrika}, 93:\penalty0 989--995, 2008.

\bibitem[Hemri et~al.(2014)Hemri, Scheuerer, Pappenberger, Bogner, and
  Haiden]{Hemri&2014}
S.~Hemri, M.~Scheuerer, F.~Pappenberger, K.~Bogner, and T.~Haiden.
\newblock Trends in the predictive performance of raw ensemble weather
  forecasts.
\newblock \emph{Geophysical Research Letters}, 41:\penalty0 9197--9205, 2014.

\bibitem[Hering and Genton(2011)]{HeringGenton2011}
A.~S. Hering and M.~G. Genton.
\newblock Comparing spatial predictions.
\newblock \emph{Technometrics}, 53:\penalty0 414--425, 2011.

\bibitem[Kleiber et~al.(2011{\natexlab{a}})Kleiber, Raftery, Baars, Gneiting,
  Mass, and Grimit]{Kleiber&a2011}
W.~Kleiber, A.~E. Raftery, J.~Baars, T.~Gneiting, C.~Mass, and E.~P. Grimit.
\newblock Locally calibrated probabilistic temperature foreasting using
  geostatistical model averaging and local {B}ayesian model averaging.
\newblock \emph{Monthly Weather Review}, 139:\penalty0 2630--2649,
  2011{\natexlab{a}}.

\bibitem[Kleiber et~al.(2011{\natexlab{b}})Kleiber, Raftery, and
  Gneiting]{Kleiber&b2011}
W.~Kleiber, A.~E. Raftery, and T.~Gneiting.
\newblock Geostatistical model averaging for locally calibrated probabilistic
  quantitative precipitation forecasting.
\newblock \emph{Journal of the American Statistical Association}, 106:\penalty0
  1291--1303, 2011{\natexlab{b}}.

\bibitem[Leutbecher and Palmer(2008)]{LeutbecherPalmer2008}
M.~Leutbecher and T.~N. Palmer.
\newblock {Ensemble forecasting}.
\newblock \emph{Journal of Computational Physics}, 227:\penalty0 3515--3539,
  2008.

\bibitem[Lindgren and Rue(2015)]{LindRue2015}
F.~Lindgren and H.~Rue.
\newblock Bayesian spatial modelling with {R-INLA}.
\newblock \emph{Journal of Statistical Software}, 63, 2015.

\bibitem[Lindgren et~al.(2011)Lindgren, Rue, and Lindstr\"{o}m]{LindRue}
F.~Lindgren, H.~Rue, and J.~Lindstr\"{o}m.
\newblock {A}n explicit link between {G}aussian fields and {G}aussian {M}arkov
  random fields: The stochastic partial differential equation approach (with
  discussion).
\newblock \emph{Journal of the Royal Statistical Society Series B},
  73:\penalty0 423--498, 2011.

\bibitem[Mass et~al.(2008)Mass, Baars, Wedam, Grimit, and Steed]{MassETAL2008}
C.~F. Mass, J.~Baars, G.~Wedam, E.~Grimit, and R.~Steed.
\newblock Removal of systematic model bias on a grid.
\newblock \emph{Weather and Forecasting}, 23:\penalty0 438--459, 2008.

\bibitem[Mat\'{e}rn(1986)]{Matern}
B.~Mat\'{e}rn.
\newblock \emph{Spatial {V}ariation}.
\newblock Springer, 2nd edition, 1986.

\bibitem[M\"{o}ller et~al.(2013)M\"{o}ller, Lenkoski, and
  Thorarinsdottir]{MoellerLenkTho2013}
A.~M\"{o}ller, A.~Lenkoski, and T.~L. Thorarinsdottir.
\newblock Multivariate probabilistic forecasting using {B}ayesian model
  averaging and copulas.
\newblock \emph{Quarterly Journal of the Royal Meteorological Society},
  139:\penalty0 982--991, 2013.

\bibitem[Molteni et~al.(1996)Molteni, Buizza, Palmer, and
  Petroliagis]{Molteni&1996}
F.~Molteni, R.~Buizza, T.~N. Palmer, and T.~Petroliagis.
\newblock The new {ECMWF} ensemble prediction system: {M}ethodology and
  validation.
\newblock \emph{Quarterly Journal of the Royal Meteorological Society},
  122:\penalty0 73--119, 1996.

\bibitem[{R Core Team}(2013)]{R2013}
{R Core Team}.
\newblock \emph{R: A Language and Environment for Statistical Computing}.
\newblock R Foundation for Statistical Computing, Vienna, Austria, 2013.
\newblock URL \url{http://www.R-project.org/}.

\bibitem[Raftery et~al.(2005)Raftery, Gneiting, Balabdaoui, and
  Polakowski]{Raftery&2005}
A.~E. Raftery, T.~Gneiting, F.~Balabdaoui, and M.~Polakowski.
\newblock Using {B}ayesian model averaging to calibrate forecast ensembles.
\newblock \emph{Monthly Weather Review}, 133:\penalty0 1155--1174, 2005.

\bibitem[Rue and Held(2005)]{RueHeld2005}
H.~Rue and L.~Held.
\newblock \emph{Gaussian {M}arkov {R}andom {F}ields. {T}heory and
  {A}pplications.}
\newblock Chapman and Hall/CRC, 2005.

\bibitem[Rue et~al.(2009)Rue, Martino, and Chopin]{RueMartChop}
H.~Rue, S.~Martino, and N.~Chopin.
\newblock {A}pproximate {B}ayesian inference for latent {G}aussian models by
  using integrated nested {L}aplace approximation (with discussion).
\newblock \emph{Journal of the Royal Statistical Society Series B},
  71:\penalty0 319--392, 2009.

\bibitem[Schefzik(2015)]{Schefzik2015}
R.~Schefzik.
\newblock Multivariate discrete copulas, with applications in probabilistic
  weather forecasting.
\newblock \emph{Publications de l'Institut de Statistique de l'Universit\'e de
  Paris}, 59:\penalty0 87--116, 2015.

\bibitem[Schefzik et~al.(2013)Schefzik, Thorarinsdottir, and
  Gneiting]{Schefzik&2013}
R.~Schefzik, T.~L. Thorarinsdottir, and T.~Gneiting.
\newblock Uncertainty quantification in complex simulation models using
  ensemble copula coupling.
\newblock \emph{Statistical Science}, 28:\penalty0 616--640, 2013.

\bibitem[Scheuerer(2014)]{Scheuerer2014}
M.~Scheuerer.
\newblock Probabilistic quantitative precipitation forecasting using ensemble
  model output statistics.
\newblock \emph{Quarterly Journal of the Royal Meteorological Society},
  140:\penalty0 1086--1096, 2014.

\bibitem[Scheuerer and B\"{u}ermann(2014)]{ScheuererBuermann2014}
M.~Scheuerer and L.~B\"{u}ermann.
\newblock Spatially adaptive post-processing of ensemble forecasts for
  temperature.
\newblock \emph{Journal of the Royal Statistical Society Series C},
  63:\penalty0 405--422, 2014.

\bibitem[Scheuerer and K\"{o}nig(2014)]{ScheuererKoenig2014}
M.~Scheuerer and G.~K\"{o}nig.
\newblock Gridded, locally calibrated, probabilistic temperature forecasts
  based on ensemble model output statistics.
\newblock \emph{Quarterly Journal of the Royal Meteorological Society},
  140:\penalty0 2582--2590, 2014.

\bibitem[Thorarinsdottir and Gneiting(2010)]{ThorarinsdottirGneiting2010}
T.~L. Thorarinsdottir and T.~Gneiting.
\newblock Probabilistic forecasts of wind speed: {E}nsemble model output
  statistics using heteroskedastic censored regression.
\newblock \emph{Journal of the Royal Statistical Society Series A},
  173:\penalty0 371--388, 2010.

\bibitem[Vrugt et~al.(2008)Vrugt, Diks, and Clark]{VrugtDiksClark2008}
J.~A. Vrugt, C.~G.~H. Diks, and M.~P. Clark.
\newblock Ensemble {B}ayesian model averaging using {M}arkov chain {M}onte
  {C}arlo sampling.
\newblock \emph{Environmental Fluid Mechanics}, 8:\penalty0 579--595, 2008.

\bibitem[Whitaker and Loughe(1998)]{WhitakerLoughe}
J.~S. Whitaker and A.~F. Loughe.
\newblock The relationship between ensemble spread and ensemble mean skill.
\newblock \emph{Monthly Weather Review}, 126:\penalty0 3292--3302, 1998.

\bibitem[Wilks(2011)]{Wilks2011}
D.~S. Wilks.
\newblock \emph{Statistical {M}ethods in the {A}tmospheric {S}ciences}.
\newblock Elsevier Academic Press, 3rd edition, 2011.

\bibitem[Wilks(2015)]{Wilks2015}
D.~S. Wilks.
\newblock Multivariate ensemble model output statistics using empirical
  copulas.
\newblock \emph{Quarterly Journal of the Royal Meteorological Society},
  141:\penalty0 945--952, 2015.

\bibitem[Wilks and Hamill(2007)]{WilksHamill2007}
D.~S. Wilks and T.~M. Hamill.
\newblock Comparison of ensemble-{MOS} methods using {GFS} reforecasts.
\newblock \emph{Monthly Weather Review}, 135:\penalty0 2379--2390, 2007.

\end{thebibliography}

\end{document}